\documentclass[twocolumn,amsmath,amssymb]{revtex4}

\usepackage{graphicx} 
\usepackage{dcolumn}  
\usepackage{bm}       

\begin{document}

\title{Renormalization of the quasiparticle hopping integrals
by spin interactions in layered copper oxides} 

\author{L. Hozoi and S. Nishimoto}
\affiliation{Max-Planck-Institut f\"{u}r Physik komplexer Systeme,
             N\"{o}thnitzer Str. 38, 01187 Dresden, Germany}

\author{C. de Graaf}
\affiliation{ICREA Research Professor at the Department of Physical 
             and Inorganic Chemistry, Universitat Rovira i Virgili, 
             Marcel$\cdot{l}$\'{i} Domingo s/n, 43007 Tarragona,
             Spain}

\date{\today}

\begin{abstract}
Holes doped within the square CuO$_2$ network specific to the cuprate
superconducting materials have oxygen $2p$ character.
We investigate the basic properties of such oxygen holes by 
wavefunction-based quantum chemical calculations on large embedded
clusters. 
We find that a $2p$ hole induces ferromagnetic correlations
among the nearest-neighbor Cu $3d$ spins.
When moving through the antiferromagnetic background the hole
must bring along this spin polarization cloud at nearby Cu
sites, which gives rise to a substantial reduction of the effective
hopping parameters.
Such interactions can explain the relatively low values inferred for
the effective hoppings by fitting the angle-resolved photoemission
data.
The effect of the background antiferromagnetic couplings of
renormalizing the effective nearest-neighbor hopping is also confirmed
by density-matrix renormalization-group model Hamiltonian
calculations for chains and ladders of CuO$_4$ plaquettes.
\end{abstract}


\maketitle

\section{Introduction}

Most theoretical models for high-temperature superconductivity are based
on Hubbard or $t\!-\!J$ Hamiltonians with one orbital per CuO$_4$
plaquette. 
For the square lattice, the tight-binding part of such models reads in 
the $\mathbf{k}$ representation
$\epsilon(\mathbf{k}) = -2t\,(\cos k_x + \cos k_y) + 4t'\cos k_x \cos k_y
                        -2t''(\cos 2k_x + \cos 2k_y) + ...$\,, where
$t$ is the hopping integral between nearest neighbors and $t'$ and $t''$ 
are hoppings between second and third order neighbors. 
The values of these parameters were estimated by fitting the angle-resolved
photoemission spectroscopy (ARPES) experimental data, see for example 
Refs.\,\cite{CuO_norman_00,CuO_schabel_98,CuO_manske_01}, and also by first
principles investigations, both periodic density-functional calculations 
within the local density approximation (LDA) \cite{CuO_hybertsen_90,CuO_oka_95,CuO_oka_01}
and embedded cluster, wavefunction-based calculations
\cite{CuO_calzado_00,CuO_calzado_01,CuO_munoz_02}. 
The analysis of the LDA conduction bands in several copper oxide compounds
yields a value of 0.4--0.5\,eV for $t$ and a ratio between the
nearest-neighbor and next-nearest-neighbor hopping integrals
$t'/t\!\approx\!0.15$ for La$_2$CuO$_4$ \cite{CuO_hybertsen_90,CuO_oka_01}
and 0.33 for Tl$_2$Ba$_2$CuO$_6$ \cite{CuO_oka_01}.
Wavefunction-based, quantum chemical calculations on finite clusters 
predict very similar nearest-neighbor hoppings, with $t$ varying between
0.5 and 0.6 eV for different cuprate superconductors \cite{CuO_calzado_00,CuO_calzado_01,CuO_munoz_02},
with the observation that $t$ correponds in this case to the propagation
of an oxygen $2p$ hole and not to $d$-like conduction-band states.
However, these estimates are quite far from the numbers obtained by 
fitting the ARPES data. Such procedures give usually nearest-neighbor
hopping integrals in the range of 0.10--0.30\,eV for hole doping and
0.10--0.20\,eV for the electron doped materials
\cite{CuO_norman_00,CuO_schabel_98,CuO_manske_01}.

This paper presents a careful analysis of the characteristics of 
oxygen $2p$ holes doped into the spin-1/2 antiferromagnetic
CuO$_2$ layer \cite{note_RefUndConf}.
We perform \textit{ab initio} multiconfiguration calculations
\cite{book_qc} on clusters that are large enough to account for charge 
\textit{and} spin relaxation and polarization effects in the near
surrounding.
We show that the hopping of the doped particle is strongly renormalized 
by spin interactions with and among the $S\!=\!1/2$ Cu $d^9$ neighbors
and that by properly taking into account these effects, good agreement is
found with the estimates extracted from the photoemission data for the
hopping integrals.
Our study is able to explain thus the difference between these ``fitted'' 
tight-binding parameters and the results of previous embedded cluster
quantum chemical investigations or periodic density-functional
calculations.
The strong renormalization of the nearest-neighbor effective hopping
found by \textit{ab initio} wavefunction-based quantum chemical methods
is confirmed by density-matrix renormalization-group (DMRG) 
\cite{dmrg_white_93,dmrg_review_scholl} model Hamiltonian calculations
on chains and ladders of CuO$_4$ plaquettes.

\section{Charge and spin configuration around a doped hole}

To illustrate the nature of an oxygen hole in the CuO$_2$ plane, we
first discuss results of multiconfiguration self consistent field
(MCSCF) calculations on a 9-plaquette square cluster, see Fig.1.
This cluster includes nine Cu ions and twenty four in-plane oxygens 
but no apex ligands. The geometrical structure corresponds to that of
the hole doped La$_{1.85}$Sr$_{1.85}$CuO$_4$ compound
\cite{LaCuO_cava87}.
We use effective core potentials (ECPs) for the $1s$,...$3s$ inner
shells of each transition metal center and the $1s$ core of the oxygens
at the boundaries of the cluster, as developed by Seijo \textit{et al.}
\cite{ecp_seijo_89} and Bergner \textit{et al.} \cite{ecp_dolg_93},
respectively.
For the outer shells of these ions we apply the following 
gaussian-type basis sets: Cu ($9s6p6d$)/[$3s3p3d$] \cite{ecp_seijo_89}
and O ($4s5p$)/[$2s3p$] \cite{ecp_dolg_93}.
Still, the rest of the ligands, i.e. those bridging the cluster
Cu ions (see Fig.1(a)), are represented with all-electron basis sets 
with a ($14s9p$)/[$4s3p$] contraction scheme \cite{ANOs_O}.
Our clusters are always embedded in large arrays of point 
charges at the experimental lattice positions that reproduce the 
crystal Madelung field.
The Cu$^{2+}$ and La$^{3+}$ nearby neighbors are represented by total 
ion potentials \cite{TIPs_CuSr}. 
The calculations were performed with the \textsc{molcas\,6} program 
package \cite{molcas6}.

Multiconfiguration, complete active space (CAS) \cite{book_qc}
calculations were carried out for a single hole ``doped'' into
the 9-plaquette cluster.
We employed a minimal CAS for constructing the multiconfiguration
wavefunction, with ten electrons distributed in all possible ways
over ten active orbitals.
Those are the nine Cu $3d_{x^2-y^2}$ orbitals plus one O $2p$ orbital.
The rest of the Cu $3d$ and O $2p$ orbitals and the other lower-energy 
orbitals are doubly occupied in all configurations and do not
participate to the construction of the MC expansion.
The ground-state has overall singlet spin multiplicity and corresponds
to the formation of a $pd$ Zhang-Rice (ZR) \cite{ZR_88} -like 
configuration on the central Cu$_c$O$_4$ plaquette.
Mulliken charge and spin populations, MCPs and MSPs, respectively,
are collected for this state in Table\,I.
The bonding (B) and antibonding (AB) $pd$ ZR orbitals are plotted in 
Fig.1(b-c).
Their occupation numbers are 1.80 and 0.20, respectively.
The Mulliken charges of the oxygen $2p_x$ and $2p_y$ atomic 
orbitals (AOs) forming the $pd$ B and AB combinations
on the ZR plaquette are similar to the values reported in
\cite{CuO_HNY_05,CuO_HN_06}. The Mulliken charges  
associated with the Cu $3d_{x^2-y^2}$ functions are slightly larger
due to the different basis sets employed here.

\begin{figure}[!t]
\includegraphics*[angle=270,width=1.0\columnwidth]{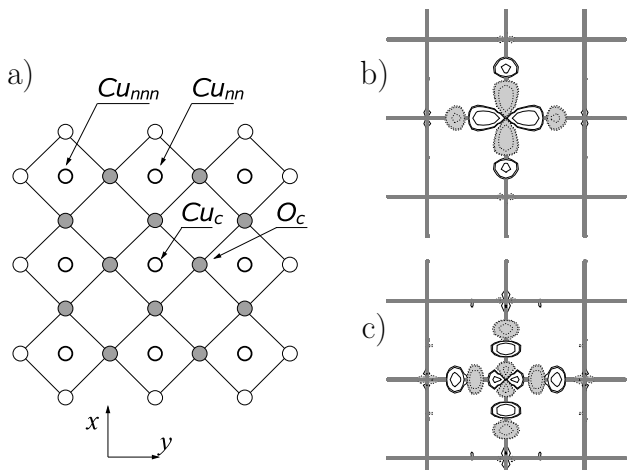}
\caption{(a) The 9-plaquette [Cu$_9$O$_{24}$] cluster employed for
studying the distribution of an O $2p$ hole.
O ions that are modeled with all-electron basis sets are shown in 
grey.
(b)-(c) B and AB $pd$ orbitals defining the ZR-like
state on the central Cu$_c$O$_4$ plaquette.
Their occupation numbers are 1.80 and 0.20.
Positive and negative lobes of the $p$ and $d$ functions are shown
in white and gray, respectively.
For the B combination the $p$ and $d$ AOs are strongly overlapping.
}
\end{figure}

The formation of CuO$_4$ $pd$ singlet states as predicted by Zhang 
and Rice \cite{ZR_88} was previously confirmed by the \textit{ab
initio} study of Calzado and Malrieu \cite{CuO_calzado_00,CuO_calzado_01}.
Two of us investigated in Refs.\,\cite{CuO_HNY_05,CuO_HN_06} 
electron-lattice interactions associated with such configurations. 
It was found that the ZR state is stabilized by lattice relaxation
effects involving a shortening by few percents of the Cu--O bonds. 
The magnitude of the stabilization energy indicates that the doped
holes form small polarons in cuprates, at least at low concentrations.
In this paper, however, we are mainly concerned with pure electronic
effects, that determine the characteristics of the ZR-like 
quasiparticle. 
All calculations are therefore performed on high-symmetry,
undistorted clusters using the average Cu--O distances reported in 
Ref.\,\cite{LaCuO_cava87}. 
For the [Cu$_9$O$_{24}$] square cluster depicted in Fig.1, both
ZR-like states and broken-symmetry solutions with the $2p$ hole
having the largest weight at a single oxygen site were obtained.
Such O $2p^5$-like states were discussed in Ref.\,\cite{CuO_HNY_05}.
The energy difference between the ZR Cu$_c$O$_4$ state and a
broken-symmetry O$_c$ $2p^5$ -like state is 5 meV in the
[Cu$_9$O$_{24}$] cluster, with the ZR configuration lower in
energy.
The broken-symmetry configurations may be favored, however, by
inter-carrier Coulomb and spin interactions.
Effects of such interactions between two or more holes were
investigated in Ref.\,\cite{CuO_LRI_06} by calculations on 
7-plaquette linear clusters.
For an undistorted 7-plaquette cluster and a single doped
hole, the energy difference between the ZR state and a
broken-symmetry $2p^5$-like state, both involving oxygens ions on
the same plaquette, is again very small, about 5 meV, but the order
of the states is reversed.
This different ordering of the states is due to finite-size
effects in the linear cluster \cite{note_GS}.
To conclude this part of our discussion, the calculations on the
large 9-plaquette cluster show that an isolated O $2p$ 
hole, i.e. an oxygen hole at sufficiently low doping, should be
viewed as part of a $pd$ ZR-like state even if the local lattice
relaxation effects are neglected.
Our estimation of the nearest-neighbor and next-nearest-neighbor
hopping matrix elements will be based on such a ZR-like 
quasiparticle picture.

\begin{table}[!b]
\caption{Mulliken population analysis illustrating the distribution
of an O $2p$ hole and the nature of the Cu--Cu spin couplings, see text.
CASSCF results for a 9-plaquette cluster.
Notations as in Fig.1 are used.
For each of the O$_{c\,}$, Cu$_{nn\,}$, and Cu$_{nnn}$ ions, there are
another three equivalent sites.}
\begin{ruledtabular}
\begin{tabular}{llr}
Relevant AOs                    &MCPs                     &MSPs    \\
\colrule
Cu$_c$          $3d_{x^2-y^2}$  &$1.17$                  & $0.06$ \\
O$_{c}^{x,y}$   $2p_{x,y}$      &$1.62$\tablenotemark[1] &$-0.01$ \\
Cu$_{nn}^{x,y}$ $3d_{x^2-y^2}$  &$1.27$                  & $0.31$ \\
Cu$_{nnn}^{xy}$ $3d_{x^2-y^2}$  &$1.28$                  &$-0.32$ \\
\end{tabular}
\end{ruledtabular}
\tablenotetext[1]{The Mulliken charges of the other in-plane oxygen
atomic orbitals are not smaller than 1.8.}
\end{table}

A distinctive feature of the results listed in Table\,I is
the nature of some of the spin couplings.
The spin populations given in the last column are obtained from
the difference between the ``up'' and ``down'' spin densities. 
The very low values associated with the Cu $d_{x^2-y^2}$ and O 
$p_x$ and $p_y$ orbitals forming the linear combinations depicted
in Fig.1(b-c) indicate indeed that the dominant contribution to the
many-electron wavefunction comes from configurations where 
the $d$ and $p$ holes on the Cu$_c$O$_4$ plaquette are coupled
to a singlet.
The fact that these numbers are not exactly zero shows that  
configurations related to different and more complicated spin
coupling schemes, involving also adjacent $d$ spins, contribute 
to the MC wavefunction too. 
Such $d\!-\!p\!-\!d$ interactions in the presence of the oxygen
hole induce actually ferromagnetic spin correlations between 
the $d_{x^2-y^2}$ electron at the ``central'' Cu$_c$ site and 
the nearest-neighbor $d_{x^2-y^2}$ electrons, as illustrated 
by the Mulliken spin populations listed in Table\,I.

The dressing of an O hole by a two-dimensional ferromagnetic
spin polarization cloud was previously suggested by Hizhnyakov
and Sigmund \cite{CuO_hizhny_88} on the basis of Hubbard-like
$pd$ model Hamiltonian calculations.
Nevertheless, the dominant interactions are those between the
$d$ and $p$ electron holes on a single CuO$_4$ plaquette such
that this object may be viewed as a ZR-\textit{like}
quasiparticle.
As expected, the data from Table\,I also show that farther 
$d\!-\!d$ interactions, between first and second order Cu 
neighbors to the central plaquette are antiferromagnetic.

\begin{table}[!t]
\caption{Nearest-neighbor hopping integrals as obtained by
CASSCF/CASSI calculations on clusters of different sizes,
see text.
Mulliken charge populations for the O $2p$ AOs forming the
$\sigma$ $p_{x,y}$--\,$d_{x^2-y^2}$ bonds on a ZR plaquette
are also listed.}
\begin{ruledtabular}
\begin{tabular}{ccc}
Cluster                   &$t$\,(eV)   &O-hole AOs, MCPs:       \\
                          &
                          &\ $p_y$, \ \ \ $p'_y$, \ \ \ $p_x$ \ \\
\colrule
$[$Cu$_{6}$O$_{19}$$]$\tablenotemark[1]
                          &0.310     &1.61, 1.61, 1.60     \\
$[$Cu$_{6}$O$_{17}$$]$\tablenotemark[2]
                          &0.165     &1.61, 1.63, 1.63     \\
                          &0.160     &1.60, 1.64, 1.63     \\
$[$Cu$_{10}$O$_{29}$$]$\tablenotemark[3]
                          &0.135     &1.65, 1.61, 1.63     \\
\end{tabular}
\end{ruledtabular}
\tablenotetext[1]{Linear cluster, all-electron (AE) basis sets
(BSs) \cite{note_2xCuO4}.}
\tablenotetext[2]{3 by 2 plaquettes, see text.
Results with AE BSs, first line, and with ECPs for the ions
outside the active region, next line.}
\tablenotetext[3]{ECPs were used for part of the Cu and O ions, see
text.}
\end{table}

\section{Renormalization of the hopping integrals by spin
interactions}

The effect of spin interactions on the nearest-neighbor
and next-nearest-neighbor hopping integrals can be evaluated by
calculations on clusters of different sizes, where the O-mediated
$d\!-\!d$ magnetic couplings are included either for all or only 
part of the transition metal neighbors.
For clusters that are large enough, we can take into account
both the ferromagnetic correlations between the $d$ electron on 
the ZR plaquette and its nearest neighbors and the effect of 
the background antiferromagnetic couplings.

The clusters we employ for our investigation always include a
2-plaquette ``active'' region with one oxygen $2p$ hole plus a
variable number of neighboring (undoped) plaquettes.
The copper and oxygen ions on the two ``active'' plaquettes are
modeled by all-electron split-valence basis sets of triple-zeta 
quality \cite{note_2xCuO4}.
Depending on the size of the cluster, for the other cluster ions
we either apply the same all-electron basis sets or use effective 
potentials for the core electrons.
As mentioned above, point charges and total ion potentials are used 
for representing the rest of the crystal.
For each cluster, we first determine the two equivalent CASSCF
solutions where the $2p$ hole is localized either on the ``left''
or on the ``right'' plaquette of the active region.
Since these states are obtained by separate SCF calculations, 
they are non-orthogonal and interacting.
Non-interacting, orthogonal eigenstates can be obtained by State 
Interaction (SI, or CASSI) calculations \cite{SI_89}. 
The effective hopping integral is half of the energy separation 
between these CASSI eigenstates \cite{note_def_t}.
In terms of matrix elements between the separately optimized
CASSCF wavefunctions, it can be written as
$t_{\mathrm{eff}} = (H_{ij} - S_{ij}H_{ii})/(1-S_{ij}^2)$,
where $H_{ij}$ and $S_{ij}$ denote Hamiltonian and overlap
matrix elements between $N$-electron states $i$ and $j$,
respectively.

For all calculations we use a minimal active space, with one 
active orbital for each $d$ and $p$ hole.
The total spin multiplicity is doublet because we always consider
an even number of plaquettes (or Cu sites) and a single doped
hole.
During the CASSCF optimization for a particular $2p$-hole 
quasilocalized state not only the electronic charge is allowed to
relax but also the spin configuration of the nearby Cu $d_{x^2-y^2}$
electrons is let to readjust, see for example the discussion in
Refs.\,\cite{CuO_kane_89,CuO_horsch_90,CuO_dagotto_90,book_fulde95}.
Our approach for estimating the effective hoppings resembles in
fact an idea proposed earlier by Eder and Becker \cite{CuO_eder90},
of setting up an effective tight-binding Hamiltonian 
$H_{\mathrm{eff}}= \sum_{ij}
                   t_{\mathrm{eff}}\,(c_{i}^{\dagger}c_{j} + h.c.)$,
where $c_{i}^{\dagger}$ and $c_{i}$ consist of a sum of products
of a ``bare'' fermion operator and a number of spin flip
operators.
We note that for each hopping process within a ``rigid'' N\'{e}el
lattice the energy of the system increases by $3J/2$
\cite{book_fulde95}.
This quantity, about 0.2\,eV (the value of the antiferromagnetic
coupling constant is $120\!-\!140$ meV \cite{CuO_J_exp}), is 
comparable with the bare nearest-neighbor hopping of 
$0.4\!-\!0.5$ eV, see
Refs.\,\cite{CuO_hybertsen_90,CuO_oka_95,CuO_oka_01,CuO_calzado_00,CuO_calzado_01,CuO_munoz_02} 
and the results discussed below.
By considering also the spin polarization effect described in
the previous section, it can be argued that the magnetic energy
associated with the disordered spins created at each hopping
\cite{book_fulde95} has the same magnitude with the bare hopping
integral, i.e. the hopping of the $2p$ hole and the spin relaxation
in the immediate vicinity imply similar time scales.
Our procedure of mapping the \textit{ab initio} data onto an effective 
tight-binding Hamiltonian including also spin relaxation effects
is thus justified.  
 
$pd$ ZR-like solutions where the oxygen hole is equally distributed
over the four ligands of a given plaquette could not be obtained
for 2-plaquette and linear 4-plaquette clusters with no distortions.
For such clusters, the O hole has always the largest weight onto 
one of the ligands connecting two Cu ions.
The smallest linear cluster where the CASSCF calculations converge
to symmetric $pd$ ZR-like solutions is a 6-plaquette cluster.
Results obtained by CASSCF and CASSI calculations on the 6-plaquette
linear cluster are shown on the first line of Table\,II.
All-electron basis sets were applied in this case for each ion of
the cluster \cite{note_2xCuO4}.

The minimal-CAS SI nearest-neighbor hopping for the 6-plaquette
linear cluster is 0.310 eV.
This value decreases dramatically, however, when the cluster is
enlarged such that it includes the other four CuO$_4$ plaquettes which
are first order neighbors of the two active plaquettes.
This is a 10-plaquette cluster, schematically drawn in Fig.2(a).
In order to reduce the computational effort we used again, as in 
the 9-plaquette square cluster, the ECPs of Seijo \textit{et al.} 
\cite{ecp_seijo_89} and Bergner \textit{et al.} \cite{ecp_dolg_93}
for the copper and oxygen ions outside the active region, with
the following basis sets for the Cu $3p$,$3d$,$4s$,$4p$ and O $2s$,$2p$
outer shells: Cu ($9s6p6d$)/[$3s2p2d$] and O ($4s5p$)/[$2s3p$].
The CASSCF/SI estimate of the nearest-neighbor hopping integral
for the 10-plaquette cluster is given on the last line of Table\,II.
This value, 0.135 eV, is less than half of the estimate for the
6-plaquette linear cluster.
As mentioned above, both charge and spin relaxation are allowed
during the SCF optimization. The Mulliken spin populations for
the CASSCF solution with the O hole localized on the left-hand
plaquette of the active region are displayed in Fig.3.
It can be easily noticed that the symmetry equivalent CASSCF
ZR-like solution with the O hole on the right-hand plaquette
implies some rearengements for the spin configuration at the
neighboring Cu sites.

\begin{figure}[!t]
\includegraphics[angle=270,width=0.95\columnwidth]{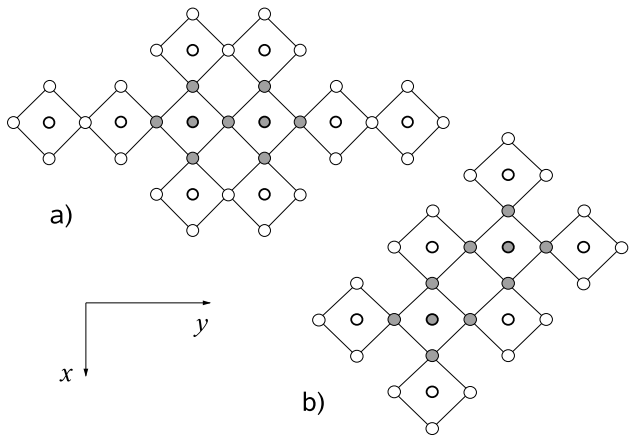}
\caption{(a) Sketch of the 10-plaquette $[$Cu$_{10}$O$_{29}$$]$ 
cluster used for evaluating the nearest-neighbor hopping $t$.
The ions in the ``active'' region are shown in grey. 
The 6-plaquette $[$Cu$_{6}$O$_{17}$$]$ cluster is obtained by
removing the four plaquettes on the same axis with the Cu
ions of the active region.
(b) The 8-plaquette $[$Cu$_{8}$O$_{24}$$]$ atomic configuration
used for evaluating $t'$.}
\end{figure}

\begin{figure}[!b]
\includegraphics[angle=270,width=0.85\columnwidth]{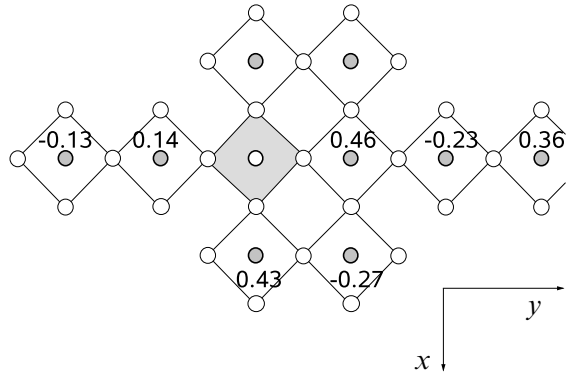}
\caption{ CASSCF Mulliken $d$-spin populations for a
10-plaquette cluster with one O hole.
The ZR-like plaquette is represented with a filled square.
The MSPs for the Cu and each O ion on this plaquette are
$0.07$ and $-0.01$, respectively.
The equivalent CASSCF state used to calculate the effective
nearest-neighbor hopping can be obtained by a reflection
through one of the symmetry planes of the cluster.
For the two CASSCF ZR-like states, different spin
populations are associated with the nearby Cu sites.}
\end{figure}

CASSCF and CASSI calculations were also carried out on a
6-plaquette cluster obtained by removing from the 10-plaquette
cluster those four plaquettes on the same axis with the copper
ions of the active region, see Fig.2(a). 
The nearest-neighbor hopping matrix element is approximately 0.17
eV in this case, see Table\,II.
This value is again much lower than the number extracted from
the linear cluster.
Still, since spin interactions are taken into account for
only part of the nearest-neighbor plaquettes around the ZR-like 
quasiparticle, it is larger than the estimate for the 10-plaquette
cluster.
It is worthwhile noting that the value of this parameter changes
only little when using effective core potentials instead
of all-electron basis sets for the ions outside the active
region, see the results for the $[$Cu$_{6}$O$_{17}$$]$ cluster
in Table\,II. 
We also point out that the occupation numbers of the $\sigma$ $p\!-\!d$
B and AB combinations on the ZR plaquettes and the
Mulliken charge populations of the $2p$ atomic orbitals accomodating
the oxygen hole are very similar for all the Cu--O clusters
discussed here. 
For comparison, Mulliken charge populations are listed in both
Table\,I and Table\,II.
Due to the symmetry properties of the atomic configurations given
in Table\,II and Fig.2(a), the $p_y$ ZR-like orbitals on a 
CuO$_4$ plaquette are not equivalent, and we denote these orbitals
as $p_y$ and $p'_y$, where $p_y$ connects the two Cu sites of 
the active region.

A different numerical experiment that confirms the effects 
illustrated in Table\,II is to replace the eight Cu$^{2+}$
$3d^9$ cations surrounding the two active plaquettes in the 
10-plaquette cluster, see Fig.2(a), with closed-shell Zn$^{2+}$
$3d^{10}$ ions \cite{note_BSs_Zn}.   
The CASSI estimate for $t$ in this $[$Cu$_{2}$Zn$_{8}$O$_{29}$$]$ 
cluster is 0.450\,eV, larger than all values listed in
Table\,II.
For a $[$Cu$_{6}$Zn$_{4}$O$_{29}$$]$ atomic configuration where
only those four Cu neighbors are replaced by Zn$^{2+}$ 
$3d^{10}$ ions which are not situated along the same line with 
the ``active'' Cu sites, $t$ is lowered to 0.330\,eV, close
to the estimate in the linear $[$Cu$_{6}$O$_{19}$$]$ cluster 
from Table\,II. 
This latter value is about two times larger than in the
$[$Cu$_{10}$O$_{29}$$]$ cluster and indicates that the effect 
of interactions with and among Cu $d$ spins which are not located
along the direction of propagation is stronger than the effect
of the collinear spin interactions.

The next-nearest-neighbor hopping, usually denoted as $t'$,
was evaluated by calculations on an 8-plaquette cluster like that
sketched in Fig.2(b).  
We applied the same ECP basis sets as for the 10-plaquette cluster
for the O and Cu ions surrounding the 2-plaquette active region,
see above.
All-electron basis sets were employed for the ligands in the 
active region \cite{note_2xCuO4}, but the two active cations
were also modeled with ECPs \cite{ecp_seijo_89}. 
The following basis sets were applied for the $3p$,$3d$,$4s$,$4p$
shells at these transition metal sites: Cu ($9s6p6d$)/[$3s3p3d$]
\cite{ecp_seijo_89}.
Our CASSI estimate for $t'$ is 0.015 eV, which gives a $t'$/$t$
ratio of 0.11 in the weakly doped La$_2$CuO$_4$ system.
A ratio $t'/t\!=\!0.12$ was obtained by fitting the ARPES data  
for an overdoped La$_{2-x}$Sr$_x$CuO$_4$ sample \cite{CuO_tohyama_99} 
and $t'/t\!=\!0.24$ for Bi$_2$Sr$_2$CaCu$_2$O$_8$
\cite{CuO_norman_00}.

A few words are in place concerning the comparison between our
present results and the results of other \textit{ab initio}
wavefunction-based investigations for copper oxide
superconductors. 
A first issue is related to the nature of the active orbital
space employed for the multiconfiguration calculations.
For the 2-plaquette ``active'' cluster region we always 
consider a CAS with three electrons and three orbitals.
These three orbitals are a bonding $p\!-\!d$ linear combination, 
an essentially nonbonding $d$ (or $d\!-\!d$) component, and an
antibonding $p\!-\!d$ orbital.
The bonding $p\!-\!d$ combination is usually
\cite{CuO_calzado_01,CuO_munoz_02} included in the inactive, 
doubly occupied orbital set, which gives an active space with 
one electron and two orbitals if other $d$ orbitals around
the 2-plaquette active cluster region are omitted.
However, the occupation numbers associated with the pair of
$p\!-\!d$ bonding and antibonding orbitals in the 3-orbital CAS,
about 1.8 and 0.2, respectively (see the discussion above and
Fig.1(b-c)), indicate significant non-dynamical correlation
effects \cite{book_qc}.
As a zeroth-order approximation, the 3-orbital CAS seems to be
then more appropriate for studying the character and the 
properties of a doped hole.
Multi-reference, difference-dedicated configuration interaction
(DDCI) \cite{ddci} calculations with reference wavefunctions
having either two or three orbitals in the active space were
performed on a hole doped 2-plaquette cluster by Calzado and
Malrieu \cite{CuO_calzado_00}.
They found that in the 2-plaquette cluster the DDCI estimate
for the nearest-neighbor hopping is in fact quite insensitive 
to the size of the active orbital space.
In addition, an iterative DDCI (IDDCI) scheme \cite{iddci} was
applied in Ref.\,\cite{CuO_calzado_00} for generating average
molecular orbitals and to obtain an unbiased representation of
the two states used to extract $t$.
It was concluded that the description of the oxygen hole in the
IDDCI wavefunction is only little affected when decreasing the 
size of the active space from three to two orbitals.

For a 2-plaquette $[$Cu$_{2}$O$_{7}$$]$ cluster and all-electron
basis sets \cite{note_2xCuO4} our CASSI estimate for the
nearest-neighbor hopping integral is 0.650\,eV, rather similar
to the IDDCI value reported in Ref.\,\cite{CuO_calzado_00},
$t\!=\!0.575$\,eV \cite{note_BSs_DDCI}.
As already mentioned above, the ``left'' and ``right'' -localized
O hole states are asymmetric in the 2-plaquette cluster.
Using the same notations as in Table\,II, the Mulliken charge
populations of the $p_y$, $p'_y$, and $p_x$ orbitals are
1.32, 1.74, and 1.68, respectively.
 
A straight comparison between our CASSCF/SI results and the
DDCI/IDDCI data of Calzado and Malrieu is not possible.
It is known \cite{CuO_calzado_00,CuO_coen_00,CuO_esther_05},
however, that DDCI/IDDCI brings rather small corrections to
the CASCI/CASSCF estimates for the nearest-neighbor hoppings
on 2-plaquette clusters in various cuprates, not larger than
$15\%$ \cite{CuO_coen_00,CuO_esther_05}.
This suggests that dynamical electron correlation is not 
essential in determining the magnitude of the hopping
integrals, as already pointed out in Refs.\,\cite{CuO_coen_00,CuO_esther_05},
for example.
In any case, \textit{the spin} polarization and relaxation 
effects at adjacent Cu sites evidenced by our calculations on 
clusters of different sizes are mainly related to non-dynamical
electron correlation, and induce changes of the hopping 
parameter about an order of magnitude larger than the corrections
brought by DDCI in 2-plaquette clusters.
We also note that with the CASSCF/SI approach we are able to
treat explicitly \textit{charge} relaxation and polarization 
due to the creation of an oxygen hole on a given CuO$_4$  
plaquette. 
In the context of \textit{ab initio} calculations for
estimating effective electronic-structure parameters such
effects are sometimes referred to as dynamical repolarization
effects.
That a compact, transparent, and rather accurate description is
obtained when the most important configurations contributing
to the many-electron wavefunction are expressed in terms of
individually optimized orbital sets was previously shown
for the calculation of magnetic exchange coupling constants 
\cite{CuO_tony95} and core-level photoionized states
\cite{3s_xps_97_02} in transition metal oxides.

An effective nearest-neighbor hopping parameter $t$ can also
be obtained by mapping the low-energy electronic states 
of the one-band Hubbard model onto those of a {\it pd} model.
The {\it pd} Hamiltonian is given by the following expression:
\begin{eqnarray*}
H= &- &t_{pd} \sum_{\langle ij\rangle\sigma}
              \left( d^\dagger_{i\sigma}p_{j\sigma} +
                     p^\dagger_{j\sigma}d_{i\sigma} \right)        \\
   &- &t_{pp} \sum_{\langle jl\rangle\sigma}
              \left(p^\dagger_{j\sigma}p_{l\sigma} +
                    p^\dagger_{l\sigma}p_{j\sigma}\right)          \\
   &+ &\epsilon_{d} \sum_{i\sigma} d^\dagger_{i\sigma} d_{i\sigma}
    +  \epsilon_{p} \sum_{j\sigma} p^\dagger_{j\sigma} p_{j\sigma} \\
   &+ &U_d \sum_i d^\dagger_{i\uparrow}d_{i\uparrow}
                  d^\dagger_{i\downarrow}d_{i\downarrow}   
     +U_p \sum_j p^\dagger_{j\uparrow}p_{j\uparrow}
                 p^\dagger_{j\downarrow}p_{j\downarrow} \ .    
\end{eqnarray*}
$t_{pd}$ is here the hopping matrix element between $p$ and
$d$ orbitals at nearest-neighbor ligand and metal sites and 
the first sum is over all pairs of such nearest neighbors.
It is assumed that $t_{pd}$ takes the same value for all 
these Cu--O pairs.
The second sum in the expression above is over all pairs of
nearest-neighbor anions and $t_{pp}$ is the corresponding
matrix element. 
We have chosen the phases of the orbitals such that the
sign of each of the two hopping matrix elements is constant.
The difference between the on-site $d$ and $p$ orbital 
energies is in the electron representation 
$\Delta_{pd} = \epsilon_d - \epsilon_p > 0$ and $U_d>0$, 
$U_p>0 $ are the on-site Coulomb repulsion energies for
Cu and O, respectively.
For the parameters of our $pd$ model we use typical values
such as $t_{pd}=1.5$, $t_{pp}=-0.8$, $\Delta_{pd}=3$, 
$U_d=8$, and $U_p=4$, in units of eV.
The effective one-band Hubbard model is given by
\begin{eqnarray*}
H= &- &t \sum_{\langle ij\rangle\sigma}
               \left( c^\dagger_{i\sigma}c_{j\sigma} +
                      c^\dagger_{j\sigma}c_{i\sigma} \right) \\
   &+ &U \sum_i c^\dagger_{i\uparrow}c_{i\uparrow}
                c^\dagger_{i\downarrow}c_{i\downarrow} \ .
\end{eqnarray*}
$t$ corresponds here to the effective hopping of the ZR-like
quasiparticle between nearest-neighbor CuO$_4$ plaquettes. 
$U>0$ is the ``on-site'' Coulomb interaction and we choose
$U\!=\!5$ eV.

Ground-state and excited-state calculations for the $pd$
and the one-band Hubbard models are performed by using 
DMRG techniques.
For hole-doped clusters of $6\!\times\!1$ and $6\!\times\!3$ 
CuO$_4$ plaquettes we found that the lowest-lying states in
the two models have the same quantum numbers and the same
character. 
We can make thus between the two models a simple one-to-one
correspondence for several low-energy states.
The effective hopping parameter $t$ is then determined such
that it minimizes a function
\begin{equation*}
\delta E =\sum_{n=1}^4 (\Delta E_n^{pd}-\Delta E_n^{\rm Hub})^2,
\end{equation*}
where $\Delta E_n^{pd}=E_n^{pd}-E_0^{pd}$ 
($\Delta E_n^{\rm Hub}=E_n^{\rm Hub}-E_0^{\rm Hub}$) are 
relative energies of the lowest four excited eigenstates 
in the $pd$ (Hubbard) model. 
The ground-state is denoted with $n\!=\!0$. 
For $6\!\times\!1$ and $6\!\times\!3$ clusters, each doped
with two holes, our mapping procedure gives effective $t$
values of $0.51$ and $0.28$ eV, respectively.
This reduction of $t$ is due not only to the inclusion of
antiferromagnetic correlations on the adjacent chains but also
to an increase of the nearest-neighbor spin correlations
along the chain.
Within the $pd$ model, the nearest-neighbor spin correlation
function changes from $-0.103$ in the $6\!\times\!1$ chain to
$-0.152$ for the central chain in the $6\!\times\!3$ 
ladder.
For more chains this quantity is only slightly modified, to
$-0.158$ for example in a $6\!\times\!5$ cluster.
The model-Hamiltonian calculations confirm thus the
renormalization effect found by \textit{ab initio} quantum
chemical methods.
If six holes are doped into the $6\!\times\!3$ cluster,
i.e. the hole concentration is kept constant with respect to
the $6\!\times\!1$ 2-hole configuration, the effective hopping
is $0.29$ eV, which shows that for this range of hole
concentrations the doping dependence is very small.


\section{Conclusions}

The interplay among charge, spin, and lattice degrees of freedom
and the formation of composite spin-lattice polarons were 
studied in cuprates (and other transition metal oxide compounds)
by many authors.
In layered copper oxides, the spin mechanism responsible for
polaronic-like behavior is associated with the fact that a charge
carrier hopping to an adjacent site must disrupt, at least at
relatively low doping, the antiferromagnetic spin configuration
in its vicinity.
Such effects were investigated by calculations based on one-band
Hubbard and $t\!-\!J$ models, see for example  
Refs.\,\cite{CuO_zhong_92,CuO_ramsak_92,CuO_mishchenko04,CuO_prelovsek06,CuO_sangiovanni06,CuO_alex_06}.
These studies show indeed that the background antiferromagnetic
correlations lead to band narrowing and a strong enhancement
of the electron-phonon couplings.
We are able to provide here \textit{ab initio} numerical
estimates for the effect of such interactions on the effective
hopping integrals.
The results can explain the relatively low values deduced for
the quasiparticle hopping parameters from fits of the
photoemission data.

Our \textit{ab initio} calculations also show that an oxygen
hole induces ferromagnetic correlations between the $d$ 
electron on the ZR-like plaquette and the nearest-neighbor
Cu $d$ spins.
Such effects are missing in one-band Hubbard or $t\!-\!J$
models.  
At this moment we are not able to quantify the strength of these
ferromagnetic interactions in the form of effective magnetic
coupling constants or spin correlation functions, at least
not at the \textit{ab initio} level.
Nevertheless, this spin polarization effect is one of the 
factors that determines the characteristics of the hole motion.
It might also offer an explanation for the disappearence of
long-range antiferromagnetic order at doping levels which are 
much lower as compared to the electron doped cuprates.
It is interesting that for an isolated chain of plaquettes,
``local'' ferromagnetic correlations in the direction of 
propagation would tend to facilitate the motion of the hole, 
although the fact that the carrier perturbs the magnetic
couplings beyond nearest neighbors results in a stronger 
and opposite effect.
In addition to that, the effective hopping parameter 
depends strongly on interactions involving spins on adjacent
chains, i.e. Cu $d$ spins which are not situated along the
direction of propagation of the hole.
Qualitatively, this is also confirmed by mapping the solutions
of a $pd$ model Hamiltonian onto those of an effective one-band
Hubbard model for clusters consisting of chains or ladders
of plaquettes. 
The inclusion of the adjacent antiferromagnetic chains in a 
three-leg ladder determines a reduction of the effective 
nearest-neighbor hopping of about fifty percent in 
comparison to the case of a single chain.



\begin{thebibliography}{9}
\bibitem{note_RefUndConf} The undoped configuration implies Cu $3d^9$
and O $2p^6$ formal valence states (each O $2p$ orbital is doubly
occupied).
\bibitem{note_GS} Finite-size effects are also responsible for a 
different order of these states in the case of CASSCF calculations
on a [Cu$_5$O$_{26}$] square cluster in Ref.\,\cite{CuO_HNY_05}.
\bibitem{note_2xCuO4} These are atomic natural orbital (ANO) basis sets
from the \textsc{molcas} library, O ($14s9p$)/[$4s3p$] and Cu 
 ($21s15p10d$)/[$5s4p3d$], see also Refs.\,\cite{ANOs_O,ANOs_Cu}.
\bibitem{note_BSs_Zn} We used effective core potentials for the innermost
12 electrons and the following ANO basis sets from the \textsc{molcas}
library for the outer shells of the Zn cations: ($9s6p5d$)/[$3s2p2d$],
see also Ref.\,\cite{ecp_seijo_89}.
\bibitem{note_BSs_DDCI} We note that different basis sets 
were used in Ref.\,\cite{CuO_calzado_00}.
\end{thebibliography}
\end{document}